\documentclass[12pt]{article}

\usepackage{graphicx}
\usepackage{epsf}

\usepackage{amsmath, amssymb}

\begin{document}
  
\begin{titlepage}

\def\thefootnote{\fnsymbol{footnote}}

\begin{center}

\hfill TU-847 \\
\hfill UT-HET 028 \\
\hfill May, 2009

\vspace{0.5cm}
{\Large\bf 
Cosmic Gamma-ray from Inverse Compton Process in
Unstable Dark Matter Scenario}

\vspace{1cm}
{\large Koji Ishiwata}$^{\it (a)}$, 
{\large Shigeki Matsumoto}$^{\it (b)}$, 
{\large Takeo Moroi}$^{\it (a)}$

\vspace{1cm}

{\it $^{(a)}${Department of Physics, Tohoku University,
    Sendai 980-8578, Japan}}

\vspace{0.5cm}

{\it $^{(b)}${Department of Physics, University of Toyama, 
    Toyama 930-8555, Japan}}

\vspace{1cm}
\abstract{ 
  
  Motivated by the PAMELA anomaly in the fluxes of cosmic-ray $e^+$
  and $e^-$, we study the cosmic $\gamma$-ray induced by the inverse
  Compton (IC) scattering process in unstable dark matter scenario
  assuming that the anomaly is due to the $e^\pm$ emission by the
  decay of dark matter.  We calculate the fluxes of IC-induced
  $\gamma$-ray produced in our Galaxy and that from cosmological
  distance, and show that both of them are significant.  If the
  $\gamma$-ray flux is precisely determined by Fermi Gamma-ray Space
  Telescope for various line-of-sight directions, it will provide an
  important test of the decaying dark matter scenario.

 }

\end{center}
\end{titlepage}

\renewcommand{\thepage}{\arabic{page}}
\setcounter{page}{1}
\renewcommand{\thefootnote}{\#\arabic{footnote}}
\setcounter{footnote}{0}

Recent observations of the fluxes of high-energy cosmic rays have made
an impact on the understanding of the nature of dark matter.  In
particular, the PAMELA experiment has observed an increasing behavior
of the positron fraction in the cosmic ray in the energy range of $10\
{\rm GeV}\lesssim E_e\lesssim 100\ {\rm GeV}$ (with $E_e$ being the
energy of $e^\pm$) \cite{Adriani:2008zr}, which cannot be explained if
we consider the conventional $e^\pm$ fluxes in astrophysics.  This
fact suggests that there may exist a non-standard source of energetic
positron (and electron) in our Galaxy.

One of the possibilities is unstable dark matter.  If dark matter has
lifetime of $O(10^{25}-10^{26}\ {\rm sec})$, and also if positron is
produced by the decay, the PAMELA anomaly may be explained.  (For
early attempts, see, for example, \cite{Huh:2008vj, Nomura:2008ru,
  Yin:2008bs, Ishiwata:2008cv, Bai:2008jt, Chen:2008md,
  Hamaguchi:2008rv, Ponton:2008zv, Ibarra:2008jk, Chen:2008qs,
  Arvanitaki:2008hq, Hamaguchi:2008ta, Chen:2008dh}.)  In addition, a
precise measurement of the total $(e^++e^-)$ flux has been performed
by the Fermi Gamma-ray Space Telescope \cite{Abdo:2009zk}, whose
results suggest that the flux of $(e^++e^-)$ is proportional to $\sim
E_e^{-3}$.  Although the interpretation of the Fermi result is still
controversial, it has been discussed that the observed spectrum may be
too hard to be consistent with the prediction of conventional
astrophysical model, and that the $e^\pm$ observed by the Fermi
experiment may be significantly contaminated by the $e^\pm$ produced
by the decay of dark matter.  Such a scenario suggests the mass of
$m_{\rm DM}\sim O(1\ {\rm TeV})$ and the lifetime of $\tau_{\rm
  DM}\sim O(10^{26}\ {\rm sec})$ \cite{Meade:2009iu, Shirai:2009fq,
  MarNomYha}.

If the decay of dark matter is the source of the extra positrons
observed by the PAMELA experiment, the emitted positron and electron
produce photon via synchrotron radiation and inverse Compton (IC)
scattering.  In our Galaxy, energy loss rates of the energetic $e^\pm$
via these processes are of the same order, but the typical energy of
the photon emitted by these processes is different.  Since the
magnetic field in our Galaxy is expected to be $O(1\ \mu{\rm G})$, the
energy of the synchrotron radiation is typically $10^{-3}\ {\rm eV}$
when the energy of $e^\pm$ is $O(1\ {\rm TeV})$.  (The synchrotron
radiation from the Galactic center is discussed in \cite{Nardi:2008ix,
  Ishiwata:2008qy}.)  On the contrary, the IC process produces
$\gamma$-rays with higher energy.  If an energetic $e^\pm$ with
$E_e\sim O(1\ {\rm TeV})$ scatters off the cosmic microwave background
(CMB) photon, $\gamma$-ray with $E_\gamma\sim O(1-10\ {\rm GeV})$ is
produced.  In addition, in our Galaxy, there exist background photons
from stars, which have higher energy than the CMB radiation.  The IC
scattering with those photons produces $\gamma$-ray with higher
energy.  Importantly, the high-energy $\gamma$-ray flux can be
precisely measured by the Fermi telescope.  Thus, in order to examine
the scenario in which the PAMELA anomaly is explained by the decay of
dark matter, it is important to study the energetic $\gamma$-ray
emitted by the IC process.

In this Letter, we study the flux of $\gamma$-ray produced by the IC
process in the decaying dark matter scenario.  We pay particular
attention to the parameter space in which the positron fraction is in
agreement with the PAMELA results.  The $\gamma$-ray emission via the
IC process in our Galaxy was also discussed in \cite{Meade:2009iu} in
connection with the recent Fermi results.  We will show that the flux
of the IC-induced $\gamma$-ray in our Galaxy may be comparable to or
larger than the expected background flux for the energy region of
$E_\gamma\sim O(1-100\ {\rm GeV})$ for some direction of line of sight
if $m_{\rm DM}\sim O(1\ {\rm TeV})$ and $\tau_{\rm DM}\sim O(10^{26}\
{\rm sec})$ to explain the PAMELA anomaly.  We will also show that the
extra-Galactic contribution is important particularly when the
$\gamma$-ray flux is observed in the direction off the Galactic
center.  Notice that the IC-induced $\gamma$-ray from the
extra-Galactic region is negligible if the PAMELA anomaly is explained
by the dark-matter annihilation \cite{Bergstrom:2008gr,
  Cirelli:2008pk, Cholis:2008qq, Feldman:2008xs, Fox:2008kb,
  Barger:2008su, Nelson:2008hj, Harnik:2008uu, Ibe:2008ye,
  Hooper:2008kv} or by the positron emission from pulsars
\cite{Hooper:2008kg}.  Thus, we propose to study the directional
dependence of the cosmic $\gamma$-ray in order to detect the signal of
IC-induced $\gamma$-ray in decaying dark matter scenario.

We first discuss the procedure to calculate the $\gamma$-ray flux.  
The total $\gamma$-ray flux is given by the sum of two contributions:
\begin{eqnarray}
  \Phi_{\gamma}^{\rm (IC)} = 
  \Phi_{\gamma}^{\rm (Galaxy)} + 
  \Phi_{\gamma}^{\rm (Cosmo)}, 
\end{eqnarray}
where the first and second terms in the right-hand side are fluxes of
$\gamma$-ray produced in our Galaxy and that from cosmological
distance (i.e., extra-Galactic contribution), respectively.  Notice
that $\Phi_{\gamma}^{\rm (Cosmo)}$ is isotropic, while
$\Phi_{\gamma}^{\rm (Galaxy)}$ depends on direction we observe.

In order to discuss the IC-induced $\gamma$-ray, it is necessary to
understand the spectrum of the parent $e^\pm$.  In our Galaxy, energetic
$e^\pm$ is approximately in a random-walk motion because of the
entangled magnetic field.  Then, the $e^{\pm}$ energy spectrum $f_e$
(i.e., number density of ($e^++e^-$) per unit energy) in our Galaxy is
described by the following diffusion equation:
\begin{eqnarray}
  K(E) \nabla^2 f_{e}(E,\vec{x})
  + \frac{\partial}{\partial E}
  \left[ b(E,\vec{x}) f_{e}(E,\vec{x}) \right]
  + Q(E,\vec{x}) = 0,
  \label{DiffEq}
\end{eqnarray}
where $K(E)$ is the diffusion coefficient, $b(E,\vec{x})$ is the
energy loss rate, and $Q(E,\vec{x})$ is the $e^{\pm}$ source term.
In considering the long-lived dark matter, the source term is given by
\begin{eqnarray}
  Q^{\rm (Galaxy)} (E_e,\vec{x})= \frac{1}{\tau_{\rm DM}}
  \frac{\rho_{\rm DM}(\vec{x})}{m_{\rm DM}}
  \frac{dN_{e}}{dE_e},
  \label{sourceterm}
\end{eqnarray}
where $\rho_{\rm DM}^{\rm (Galaxy)}$ is energy density of dark matter
and $dN_{e}/dE$ is energy distribution of $e^\pm$ from the decay of
single dark matter.  In our study, we adopt the isothermal halo
density profile \cite{IsoThermal}
\begin{eqnarray}
  \rho_{\rm DM}^{\rm (Galaxy)} (r) 
  =
  \rho_\odot \frac{r_{\rm core}^2 + r_\odot^2}{r_{\rm core}^2+r^2},
  \label{eq:isothermal}
\end{eqnarray}
where $\rho_\odot\simeq 0.43\ {\rm GeV/cm^3}$ is the local halo
density, $r_{\rm core}\simeq 2.8\ {\rm kpc}$ is the core radius,
$r_\odot\simeq 8.5\ {\rm kpc}$ is the distance between the Galactic
center and the solar system, and $r$ is the distance from the Galactic
center.  In studying the propagation of the cosmic-ray $e^\pm$, we
assume some of the parameters suggested by the so-called MED
propagation model \cite{Delahaye:2007fr}; we use $K(E)=0.0112\ {\rm
  kpc^2/Myr}\times (E_e/1\ {\rm GeV})^{0.70}$, while the diffusion
zone is approximated by a cylinder with the half-height of $L=4\ {\rm
  kpc}$ and the radius of $R=20\ {\rm kpc}$.  In addition, the energy
loss rate $b$ is given by the sum of the contributions from the
synchrotron-radiation and the IC processes: $b^{\rm
  (Galaxy)}(E_e,\vec{x})=b_{\rm synch}^{\rm
  (Galaxy)}(E_e,\vec{x})+b_{\rm IC}^{\rm (Galaxy)}(E_e,\vec{x})$.  For
the calculation of $b_{\rm synch}^{\rm (Galaxy)}$, we approximate that
the strength of the magnetic flux density $B$ is independent of
position in our Galaxy; then we obtain
\begin{eqnarray}
  b_{\rm synch}^{\rm (Galaxy)}(E_e) = 
   \sigma_{\rm T} \gamma_e^2 B^2,
\end{eqnarray}
with $\gamma_e=E_e/m_e$ and $\sigma_{\rm T}$ being the cross section of
the Thomson scattering.  In our numerical study, we use $B=3\ \mu{\rm
G}$. Furthermore, $b_{\rm IC}^{\rm (Galaxy)}$ is given by
\begin{eqnarray}
  b_{\rm IC}^{\rm (Galaxy)} (E_e, \vec{x}) = 
  \int dE_{\gamma} dE_{\gamma_{\rm BG}} 
  (E_{\gamma}-E_{\gamma_{\rm BG}})
  \frac{d\sigma_{\rm IC}}{dE_\gamma}
  f_{\gamma_{\rm BG}} (E_{\gamma_{\rm BG}}, \vec{x}),
\end{eqnarray}
where the differential cross section for the IC process is expressed
as \cite{Blumenthal:1970gc}
\begin{eqnarray}
  \frac{d\sigma_{\rm IC}}{dE_\gamma}
   =
  \frac{3\sigma_{\rm T}}{4\gamma_e^2 E_{\gamma_{\rm BG}}}
  \left[
    2 q \ln q + (1 + 2q) (1 - q) 
    + \frac{(\Gamma_e q)^2 (1 - q)}{2 (1 + \Gamma_e q)}
  \right],
\end{eqnarray}
with $\Gamma_e=4\gamma_eE_{\gamma_{\rm BG}}/m_e$, $q=E_\gamma/\Gamma_e
(E_e - E_\gamma)$, and $f_{\gamma_{\rm BG}}$ is the spectrum of the
background photon.  Kinematically, $1/4\gamma_e^2\leq q\leq 1$ is
allowed.  The background photon in our Galaxy has three components: (i)
star light concentrated in the Galaxy, (ii) star light re-scattered by
dust, and (iii) the CMB radiation.  The spectrum of the CMB radiation
is isotropic and well known, while those of the first and second
components depend on the position.  We use the data of interstellar
radiation field provided by the GALPROP collaboration \cite{Galprop},
which is based on \cite{Porter:2005qx}, to calculate $f_{\gamma_{\rm
    BG}}$ in our Galaxy.

In the present choice of the propagation model, the typical
propagation length of electron per time scale of the energy loss is
estimated to be $O(0.1\ {\rm kpc})$ for $E_e\sim 100\ {\rm GeV}$ (and
it becomes shorter as the energy increases).  In such a case, the
$e^\pm$ spectrum at the position $\vec{x}$ is well approximated by
\begin{eqnarray}
  f_e^{\rm (Galaxy)} (E_e,\vec{x})
  = \frac{1}{b^{\rm (Galaxy)} (E_e,{\vec{x}})}
  \frac{\rho_{\rm DM}^{\rm (Galaxy)} (\vec{x})}{\tau_{\rm DM} m_{\rm DM}}
  \int_{E_e}^{\infty} dE_e^{\prime} \frac{dN_e}{dE_e^{\prime}}.
  \label{f_e(Gal)}
\end{eqnarray}
In our numerical analysis, we adopt the above approximated formula for
the $e^\pm$ spectrum in our Galaxy.  Then, $\gamma$-ray flux from
direction ($b,l$), where $b$ and $l$ are Galactic latitude and
longitude, respectively, is obtained by line-of-sight (l.o.s) integral
of $\gamma$-ray energy density per unit time and unit energy as
\begin{eqnarray}
  \Phi_{\gamma}^{\rm (Galaxy)} (b,l)=
  \frac{1}{4 \pi} \int_{\rm l.o.s} d \vec{l} 
  L_{\rm IC}(E_{\gamma}, \vec{l}),
\end{eqnarray}
where
\begin{eqnarray}
  L_{\rm IC} (E_\gamma, \vec{x})
  = \int d E_e d E_{\gamma_{\rm BG}} 
  \frac{d\sigma_{\rm IC}}{dE_\gamma}
  f_{\gamma_{\rm BG}} (E_{\gamma_{\rm BG}}, \vec{x})
    f_e^{\rm (Galaxy)} (E_e, \vec{x}).
\end{eqnarray}

In studying the $\gamma$-ray from cosmological distance, we need to
understand the $e^\pm$ spectrum in the extra-Galactic region.  In such
a region, the $e^\pm$ spectrum is independent of the position, so the
spectrum should obey
\begin{eqnarray}
  \frac{\partial f_e(t, E_e)}{\partial t} =
  H E_e \frac{\partial f_e(t, E_e)}{\partial E_e}
  + \frac{\partial}{\partial E_e}
  \left[ b(t, E_e) f_{e}(t, E_e) \right]
  + Q(t, E_e),
  \label{Boltzmann_C}
\end{eqnarray}
where $H$ is the expansion rate of the universe.  Contrary to the case
in our Galaxy, only the IC process with the CMB radiation contributes to
the energy-loss process.  Since the typical energy of the CMB radiation
is so low that $e^\pm$ becomes non-relativistic in the center-of-mass
energy of the IC process.  Then, taking into account the red-shift of
the CMB radiation, the energy loss rate is given by
\begin{eqnarray}
  b^{\rm (Cosmo)} (t, E_e) = 
  \frac{4}{3} \sigma_{\rm T} \gamma_e^2 \rho_{\rm CMB}^{\rm (now)} 
  (1+z)^4,
\end{eqnarray}
where $\rho_{\rm CMB}^{({\rm now})}\simeq 0.26\ {\rm eV/cm^3}$ is the
present energy density of the CMB, and $z$ is the red-shift.

The typical time scale of the energy loss due to the IC process is
estimated to be $E_e/b^{\rm (Cosmo)}$, and is of the order of $10^{14}\
{\rm sec}$ for $E_e=100\ {\rm GeV}$ (and becomes shorter as $E_e$
increases).  Because the energetic $e^\pm$ loses its energy via the IC
process before the energy is red-shifted, we neglect the terms of $O(H)$
in Eq.\ \eqref{Boltzmann_C} and obtain
\begin{eqnarray}
  f_e^{({\rm Cosmo})} (t, E_e)
  = \frac{1}{b^{({\rm Cosmo})}(t, E_e)}
  \frac{\rho_{\rm DM}^{({\rm now})}(1+z)^3}{\tau_{\rm DM} m_{\rm DM}}
  \int_{E_e}^{\infty} dE_e^{\prime} \frac{dN_e}{dE_e^{\prime}},
  \label{f_e(Cosmo)}
\end{eqnarray}
where $\rho_{\rm DM}^{({\rm now})}\simeq 1.2\times 10^{-6}\ {\rm
  GeV}/{\rm cm}^3$ is the present energy density of dark matter.
Then, taking into account red-shift of scattered photon spectrum
and the dilution due to the expansion of the universe, we obtain
the $\gamma$-ray from cosmological distance as
\begin{eqnarray}
  \Phi_{\gamma}^{\rm (Cosmo)} = 
  \frac{1}{4 \pi} \int dt \frac{1}{(1+z)^3} L_{\rm IC}(t,E_{\gamma}),
\end{eqnarray}
where
\begin{eqnarray}
  L_{\rm IC}(t,E_{\gamma})
  = 
  (1+z) \int d E_e d E_{\gamma_{\rm BG}} 
  \left[
  \frac{d\sigma_{\rm IC}}{dE'_\gamma}
  \right]_{E'_\gamma=(1+z)E_\gamma}
  f_{\gamma_{\rm BG}}^{\rm (CMB)} (t, E_{\gamma_{\rm BG}})
  f_e^{\rm (Cosmo)} (t, E_e),
   \nonumber \\
\end{eqnarray}
with $f_{\gamma_{\rm BG}}^{\rm (CMB)} (t, E_{\gamma_{\rm BG}})$ being
the spectrum of the CMB radiation at the time $t$.  Notice that the
astrophysical uncertainty is small in the extra-Galactic contribution,
as is obvious from the above expression.

Now, we are at the position to show our numerical results.  To make
our discussion simple, we concentrate on the case where dark matter
dominantly decays into $\mu^+\mu^-$ pair as an example.  In such a
case, energetic $e^\pm$ are produced by the decay of $\mu^\pm$.  Then,
the positron fraction can be in good agreement with the PAMELA result
while the total $(e^++e^-)$ flux can be consistent with the Fermi data
if $m_{\rm DM}\sim O(1\ {\rm TeV})$ and $\tau_{\rm DM}\sim O(10^{26}\
{\rm sec})$ \cite{Meade:2009iu}.  Here, we use the following four
sample points: $(m_{\rm DM},\tau_{\rm DM})=(1\ {\rm TeV}, 4.5\times
10^{26}\ {\rm sec})$, $(2\ {\rm TeV}, 2.2\times 10^{26}\ {\rm sec})$,
$(4\ {\rm TeV}, 1.1\times 10^{26}\ {\rm sec})$, and $(10\ {\rm TeV},
1.0\times 10^{26}\ {\rm sec})$.  With these parameters, the $e^+$ and
$e^-$ fluxes become consistent with the PAMELA and the Fermi data by
relevantly choosing the background fluxes.  Here, we assume that the
background $e^-$ flux obeys the power law, $\Phi_{e^-}^{\rm
  (BG)}=AE_e^{\gamma}$, with $A$ and $\gamma$ being free parameters,
while the background $e^+$ flux of $\Phi_{e^+}^{\rm (BG)}=4.5(E_e/1\
{\rm GeV})^{0.7}/(1+650(E_e/1\ {\rm GeV})^{2.3}+1500(E_e/1\ {\rm
  GeV})^{4.2})\ {\rm GeV}^{-1}{\rm cm}^{-2}{\rm sec}^{-1}{\rm
  sr}^{-1}$ \cite{Baltz:1998xv} is adopted.  Then, for some choice of
$A$ and $\gamma$, the positron fraction becomes consistent with the
PAMELA result.  Meanwhile, the Fermi data suggests that
$(\Phi_{e^-}^{\rm (obs)}+\Phi_{e^+}^{\rm (obs)})$ is approximately
proportional to $E_e^{-3}$.  However, we believe that the
interpretation of the Fermi data is rather controversial because the
$(e^++e^-)$ flux is sensitive to the spectra of background $e^+$ and
$e^-$ which still have significant uncertainties.  So, we
conservatively use the Fermi data as an upper bound on the $(e^++e^-)$
flux; we checked that the total $(e^++e^-)$ fluxes in our sample
points are below the observed flux reported by the Fermi experiment:
$E_e^3(\Phi_{e^-}+\Phi_{e^+})\lesssim 150\ {\rm
  GeV^2sec^{-1}m^{-2}sr^{-1}}$ for $20\ {\rm GeV}\lesssim E_e\lesssim
1\ {\rm TeV}$ \cite{Abdo:2009zk}.

The flux of the high energy cosmic $\gamma$-ray has been recently
measured by the Fermi experiment.  The Fermi collaboration has shown
preliminary results after averaging over the direction in the region
of $0^\circ\leq l<360^\circ$ and $10^\circ\leq b\leq 20^\circ$.  Then,
they found that the averaged $\gamma$-ray flux is
$E_\gamma^2\Phi_\gamma^{\rm (obs)}\sim O(10^{-5}-10^{-6}\ {\rm GeV
  sec^{-1}cm^{-2}sr^{-1}})$ for the energy range of $0.1\ {\rm
  GeV}\lesssim E_\gamma\lesssim 10\ {\rm GeV}$ \cite{TITECH_Workshop}.
We have calculated the same averaged flux in the present case and
found that the flux from the IC process is order-of-magnitude smaller
than the Fermi data.

However, this fact does not necessarily mean that the study of the
IC-induced $\gamma$-ray has no importance.  This is because the signal
and background $\gamma$-ray fluxes are expected to have strong
directional dependence.  In particular, there exists stronger stellar
activity near the Galactic center, and hence the background
$\gamma$-ray flux is significantly reduced for the direction away from
the Galactic center.  Thus, if we focus on a particular direction in
which small background $\gamma$-ray flux is expected, a signal of the
IC-induced $\gamma$-ray may be observed.  Furthermore, the IC-induced
$\gamma$-ray flux may be enhanced for the energy region of
$E_\gamma\gtrsim 10\ {\rm GeV}$.

\begin{figure}
  \centerline{\epsfxsize=0.75\textwidth\epsfbox{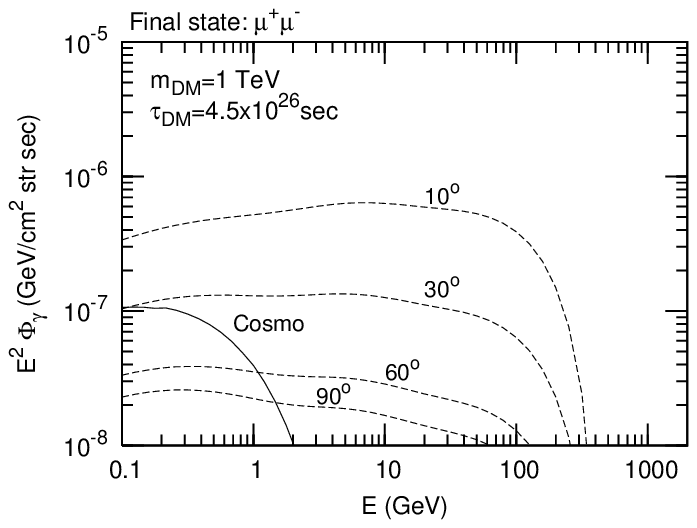}}
  \caption{\small Flux of IC-induced cosmic $\gamma$-ray for $m_{\rm
      DM}=1\ {\rm TeV}$ and $\tau_{\rm DM}=4.5\times 10^{26}\ {\rm
      sec}$.  The solid line is the flux from the cosmological
    distance, which is isotropic, while the dashed ones are Galactic
    contributions from the directions $(b,l)=(10^\circ,0^\circ)$,
    $(30^\circ,0^\circ)$, $(60^\circ,0^\circ)$, and
    $(90^\circ,0^\circ)$, from the top to the bottom. }
  \label{fig:icfluxM01TeV}
  \vspace{7mm}
  \centerline{\epsfxsize=0.75\textwidth\epsfbox{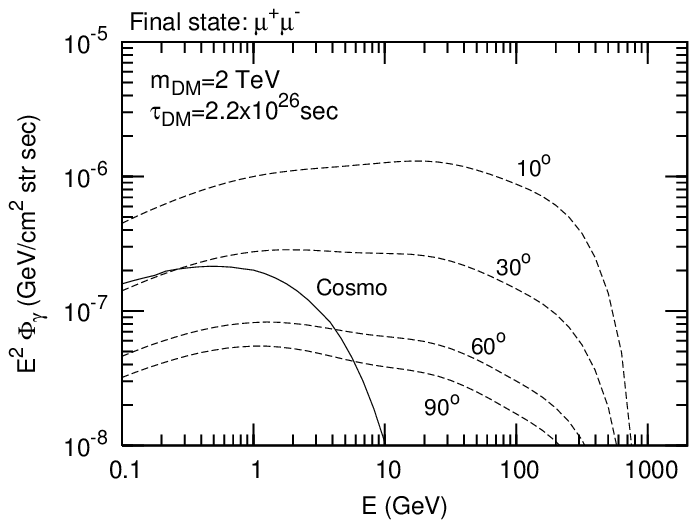}}
  \caption{\small Same as Fig.\ \ref{fig:icfluxM01TeV}, except for
    $m_{\rm DM}=2\ {\rm TeV}$ and $\tau_{\rm DM}=2.2\times 10^{26}\
    {\rm sec}$.}
  \label{fig:icfluxM02TeV}
\end{figure}

\begin{figure}
  \centerline{\epsfxsize=0.75\textwidth\epsfbox{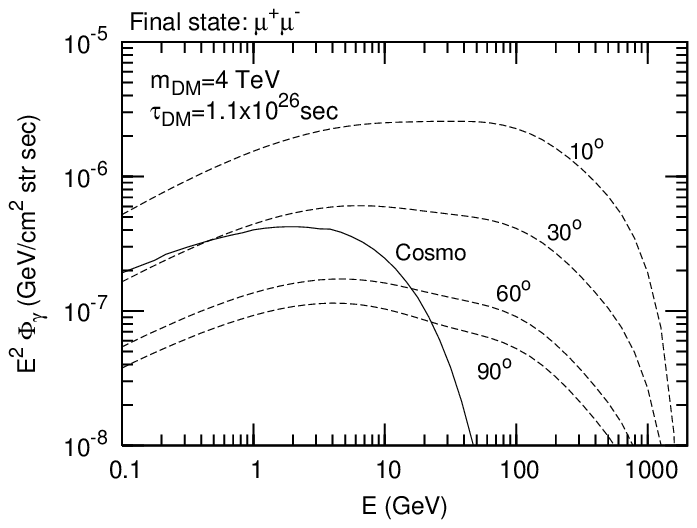}}
  \caption{\small Same as Fig.\ \ref{fig:icfluxM01TeV}, except for
    $m_{\rm DM}=4\ {\rm TeV}$ and $\tau_{\rm DM}=1.1\times 10^{26}\
    {\rm sec}$.}
  \label{fig:icfluxM04TeV}
  \vspace{7mm}
  \centerline{\epsfxsize=0.75\textwidth\epsfbox{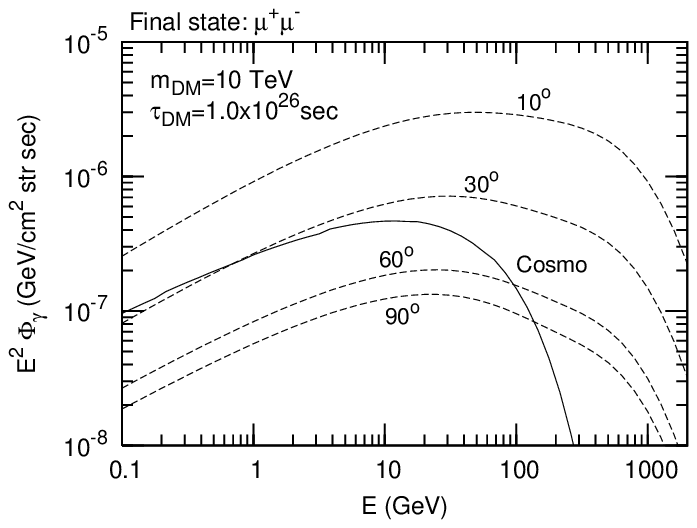}}
  \caption{\small Same as Fig.\ \ref{fig:icfluxM01TeV}, except for
    $m_{\rm DM}=10\ {\rm TeV}$ and $\tau_{\rm DM}=1.0\times 10^{26}\
    {\rm sec}$.}
  \label{fig:icfluxM10TeV}
\end{figure}

In Figs.\ \ref{fig:icfluxM01TeV} $-$ \ref{fig:icfluxM10TeV}, we plot
the cosmic $\gamma$-ray flux from the IC process for $m_{\rm DM}=1\
{\rm TeV}$, $2\ {\rm TeV}$, $4\ {\rm TeV}$, and $10\ {\rm TeV}$.  In
the Figures, we show the Galactic and extra-Galactic contributions
separately.  For the Galactic contribution, we show the results for
$(b,l)=(10^\circ,0^\circ)$, $(30^\circ,0^\circ)$,
$(60^\circ,0^\circ)$, and $(90^\circ,0^\circ)$.  We have also
calculated the $\gamma$-ray flux from other direction of
line-of-sight.  For $b<10^\circ$ (with $l=0^\circ$), the results are
almost the same as those for $(b,l)=(10^\circ,0^\circ)$ irrespective
of $m_{\rm DM}$ as far as the present procedure of the calculation is
adopted.  For $l=180^\circ$, the IC-induced $\gamma$-ray flux becomes
insensitive to $b$ and is similar to that for
$(b,l)=(90^\circ,0^\circ)$.

As one can see, for the direction close to the Galactic center,
Galactic contribution is larger than the extra-Galactic one.  For
$(b,l)=(10^\circ -30^\circ,0^\circ)$, the flux can be as large as
$E_\gamma^2\Phi_\gamma^{\rm (IC)}\sim O(10^{-6}-10^{-7}\ {\rm GeV
  sec^{-1}cm^{-2}sr^{-1}})$ up to $E_\gamma\sim 10-100\ {\rm GeV}$,
depending on the mass of dark matter.  On the contrary, for the
direction away from the Galactic center, the $\gamma$-ray from the IC
process is dominated by the extra-Galactic contribution.  In such a
case, the peak of the IC-induced $\gamma$-ray spectrum is around
$E_\gamma\sim 0.2\ {\rm GeV}\times (m_{\rm DM}/1\ {\rm TeV})^2$, and
the height of the peak can be as large as $E_\gamma^2\Phi_\gamma^{\rm
  (IC)}\sim O(10^{-7}\ {\rm GeV sec^{-1}cm^{-2}sr^{-1}})$.  Thus, if
the $\gamma$-ray fluxes for various directions are determined, they
may provide significant test of the unstable dark matter scenario.

Using EGRET \cite{Sreekumar:1997un} data, information about the
directional dependence of the $\gamma$-ray flux was discussed in
\cite{Strong:2004de}.  For $E_\gamma\sim O(1\ {\rm GeV})$, the
observed flux is $E_\gamma^2\Phi_\gamma^{\rm (obs)}\sim
O(10^{-6}-10^{-7}\ {\rm GeV sec^{-1}cm^{-2}sr^{-1}})$, which is of the
same order of the IC-induced $\gamma$-ray flux for $m_{\rm DM}\sim
O(1\ {\rm TeV})$.  (The observed flux becomes smaller for higher
Galactic latitude.)  However, for such energy region, the error in the
observed flux is very large.  In addition, for directions away from
the Galactic center, no information is currently available for larger
$E_\gamma$.  So, our knowledge about the $\gamma$-ray spectrum, in
particular for directions away from the Galactic center, is still
unsatisfactory and is not enough to perform a reliable test of the
unstable dark matter scenario.  Thus, more precise determination of
the spectrum is suggested.  If detailed data about the $\gamma$-ray
spectrum becomes available for various directions by the Fermi
experiment, it will provide a significant information about the
unstable dark matter scenario.  In particular, study of the
$\gamma$-ray flux from directions away from the Galactic center is
important.  For such directions, the $\gamma$-rays from the Galactic
activities are suppressed.  Thus, the background flux is expected to
be comparable to or even smaller than the extra-Galactic contribution
of the IC-induced $\gamma$-ray flux.  Then, if the $\gamma$-ray
spectrum for such a direction is precisely determined, we may see a
signal of the IC-induced $\gamma$-ray.  Of course, the measurement of
the flux from directions close to the Galactic center, which is
sensitive to the Galactic contribution, is also important.

In this Letter, we have studied the IC-induced $\gamma$-ray, assuming
that the PAMELA anomaly in the $e^+$ fraction is due to the decay of
dark matter.  We have calculated the Galactic and extra-Galactic
contributions separately, and shown that both of them are important.
In our study, we have considered the case that dark matter dominantly
decays into $\mu^+\mu^-$ pair.  However, as far as the PAMELA anomaly
is explained in the decaying dark matter scenario, the IC-induced
$\gamma$-ray flux is expected to be of the same order irrespective of
dominant decay process.  This is because, in order to explain the
PAMELA anomaly, energetic $e^\pm$ should be produced by the decay,
which induces the IC process.  It should be also noticed that the
production rate of energetic $e^\pm$ is extremely suppressed in the
extra-Galactic region in other scenarios of explaining the PAMELA
anomaly, like the annihilating dark matter scenario and the pulser
scenario.  Therefore, the IC-induced $\gamma$-ray from cosmological
distance is negligible in those cases, and hence it is a unique signal
in the decaying dark matter scenario.  Information about the
IC-induced $\gamma$-ray may be obtained from the observations for
various directions.  In particular, once the $\gamma$-ray fluxes for
directions off the Galactic center are precisely determined, we should
carefully study the data to extract the IC-induced $\gamma$-ray from
the extra-Galactic region, which has small astrophysical uncertainty.

So far, we have concentrated on the case that the direct production of
energetic photon by the dark matter decay is negligible.  However, it
is often the case that dark matter decays into photon (and something
else), or into hadrons whose decay products include energetic photons.
If so, those photons also become the source of high energy cosmic
$\gamma$-ray and the cosmic $\gamma$-ray flux may be enhanced.  The
spectrum of the cosmic $\gamma$-ray depends on the properties of the
decaying dark matter.  Detailed discussion for several specific dark
matter models will be given elsewhere \cite{IshMatMor}.

\noindent
{\it Acknowledgments:}
This work was supported in part by Research Fellowships of the Japan
Society for the Promotion of Science for Young Scientists (K.I.), and
by the Grant-in-Aid for Scientific Research from the Ministry of
Education, Science, Sports, and Culture of Japan, No.\ 21740174 (S.M.)
and No.\ 19540255 (T.M.).

\end{document}